\documentclass[12pt]{article}
\usepackage{comment}
\usepackage{amsmath}
\usepackage{amssymb}
\usepackage{graphicx}
\usepackage{subcaption}
\usepackage{url}
\usepackage[sort&compress, numbers, merge]{natbib}
\usepackage[colorlinks=true, linkcolor=blue, citecolor=blue,
urlcolor=black]{hyperref}

\begin{document}

\baselineskip 0.6cm

\def\simgt{\mathrel{\lower2.5pt\vbox{\lineskip=0pt\baselineskip=0pt
           \hbox{$>$}\hbox{$\sim$}}}}
\def\simlt{\mathrel{\lower2.5pt\vbox{\lineskip=0pt\baselineskip=0pt
           \hbox{$<$}\hbox{$\sim$}}}}
\def\simprop{\mathrel{\lower3.0pt\vbox{\lineskip=1.0pt\baselineskip=0pt
             \hbox{$\propto$}\hbox{$\sim$}}}}
\def\tr{\mathop{\rm tr}}
\def\SU{\mathop{\rm SU}}

\begin{titlepage}

\begin{flushright}
UT-17-02 \\
KYUSHU-RCAPP 2017-01 \\
IPMU17-0016
\end{flushright}

\vskip 1.1cm

\begin{center}

{\LARGE \bf 
Cornering Compressed Gluino at the LHC
}

\vskip 1.2cm

Natsumi Nagata${}^{1}$,
Hidetoshi Otono${}^{2}$, 
and
Satoshi Shirai${}^{3}$
\vskip 0.5cm

{\it
$^1$ Department of Physics, University of Tokyo, 
Tokyo 113-0033,
Japan\\
$^2$ Research Center for Advanced Particle Physics, Kyushu University,
 Fukuoka 812-8581, Japan\\ 
$^3$ {Kavli Institute for the Physics and Mathematics of the Universe
 (WPI), \\The University of Tokyo Institutes for Advanced Study, \\ The
 University of Tokyo, Kashiwa 277-8583, Japan}
}

\vskip 1.0cm

\abstract{
 We discuss collider search strategies of gluinos which are highly
 degenerate with the lightest neutralino in mass. This scenario is
 fairly difficult to probe with conventional search strategies at
 colliders, and thus may provide a hideaway of
 supersymmetry. Moreover, such a high degeneracy plays an important role
 in dark matter physics as the relic abundance of the lightest
 neutralino is significantly reduced via coannihilation. In this paper,
 we discuss ways of uncovering this scenario with the help of longevity
 of gluinos; if the mass difference between the lightest neutralino and
 gluino is $\lesssim 100$~GeV and squarks are heavier than gluino, then
 the decay length of the gluino tends to be of the order of the
 detector-size scale. Such gluinos can be explored in the searches of
 displaced vertices, disappearing tracks, and anomalously large energy
 deposit by (meta)stable massive charged particles. We find that these
 searches are complementary to each other, and by combining their
 results we may probe a wide range of the compressed gluino region in
 the LHC experiments.  
}

\end{center}
\end{titlepage}

%%%%%%%%%%%%%%%%%%%%%%%%%%%%%%%%%%%%%%%%%%%%%%%%%%%%
\section{Introduction}
\label{sec:intro}
%%%%%%%%%%%%%%%%%%%%%%%%%%%%%%%%%%%%%%%%%%%%%%%%%%%%

Supersymmetric (SUSY) extensions of the Standard Model (SM) have been
thought of as the leading candidate of physics beyond the SM. In
particular, weak-scale SUSY has various attractive features---the
electroweak scale is stabilized against quantum corrections, gauge
coupling unification is achieved, and so on---and
therefore has widely been studied so far. 
This paradigm is, however, under strong pressure from the results
obtained at the Large Hadron Collider (LHC).
Direct searches of SUSY particles impose stringent limits on
their masses, especially those of colored particles such as squarks and
gluino \cite{ATLAS-CONF-2016-078, *ATLAS-CONF-2016-052,
*CMS-PAS-SUS-16-015}. In addition, the observed value of the mass of the
SM-like Higgs boson \cite{Aad:2012tfa, *Chatrchyan:2012ufa,
*Aad:2015zhl}, $m_h\simeq 125$~GeV, may also imply that SUSY particles
are rather heavy. In the minimal supersymmetric Standard Model (MSSM),
the tree-level value of the SM-like Higgs boson mass is smaller than the
$Z$-boson mass \cite{Inoue:1982ej, *Flores:1982pr}, and we need sizable
quantum corrections in order to explain the discrepancy between the
tree-level prediction and the observed value. It turns out that
sufficiently large radiative corrections are provided by stop-loop
diagrams \cite{Okada:1990vk,*Okada:1990gg,*Ellis:1990nz, *Haber:1990aw,
*Ellis:1991zd} if the stop masses are much larger than the electroweak
scale.

The SUSY SM with heavy SUSY particles (or, equivalently, with a high
SUSY-breaking scale) have various advantages from the
phenomenological point of view \cite{Wells:2003tf,*Wells:2004di,
ArkaniHamed:2004fb,*Giudice:2004tc,*ArkaniHamed:2004yi,
Hall:2011jd, Hall:2012zp, Ibe:2006de,Ibe:2011aa, *Ibe:2012hu,
Arvanitaki:2012ps, ArkaniHamed:2012gw, Evans:2013lpa, *Evans:2013dza,
Nomura:2014asa}; {\it i}) severe limits from the measurements of
flavor-changing processes and electric dipole moments can be evaded
\cite{Gabbiani:1996hi, Moroi:2013sfa, *McKeen:2013dma,
*Altmannshofer:2013lfa, *Fuyuto:2013gla}; {\it ii}) heavy sfermions do
not spoil successful gauge coupling unification if gauginos and
Higgsinos remain around the TeV scale \cite{Hisano:2013cqa}; {\it iii})
the dimension-five proton decay caused by the color-triplet Higgs
exchange \cite{Sakai:1981pk, *Weinberg:1981wj}, which was problematic
for weak-scale SUSY \cite{Goto:1998qg, *Murayama:2001ur}, is suppressed
by sfermion masses and thus the current proton decay bound can be evaded
if SUSY particles are heavy enough \cite{Liu:2013ula, *Hisano:2013exa,
*Dine:2013nga, *Nagata:2013sba, *Hall:2014vga, *Evans:2015bxa,
*Bajc:2015ita, *Ellis:2015rya, *Ellis:2016tjc}, making the minimal SUSY
SU(5) grand unification \cite{Sakai:1981gr, *Dimopoulos:1981zb} viable;
{\it iv}) cosmological problems in SUSY theories, such as the gravitino
problem \cite{Pagels:1981ke, *Weinberg:1982zq, *Khlopov:1984pf,
Kawasaki:2004qu, *Kawasaki:2008qe} and the Polonyi problem
\cite{Coughlan:1983ci} can be avoided. Nevertheless, high-scale SUSY
models may have a potential problem regarding dark matter. In SUSY SMs
with $R$-parity conservation, the lightest SUSY particle (LSP) is
stable and thus can be a dark matter candidate. In particular, if
the LSP is the lightest neutralino, its relic abundance is determined by the
ordinary thermal freeze-out scenario. Then, it turns out that if the
mass of the neutralino LSP is well above the weak scale, its thermal relic
abundance tends to exceed the observed value of dark matter density,
$\Omega_{\rm DM} h^2 \simeq 0.12$ \cite{Ade:2015xua}. Thus, the
requirement of $\Omega_{\rm DM} h^2 \lesssim 0.12$ imposes a severe
constraint on models with high-scale SUSY breaking.\footnote{In fact,
environmental selection with multiverse may naturally give the condition
$\Omega_{\rm DM} h^2 \lesssim 0.12$ and favor a ``Spread SUSY''-type
spectrum \cite{Hall:2011jd, Hall:2012zp, Nomura:2014asa}. }

In order to avoid over-production of the neutralino LSP in the high-scale
SUSY scenario, it is necessary to assure a large annihilation cross
section for the LSP. A simple way to do that is to assume the neutralino
LSP to be an almost pure SU(2)$_L$ multiplet, {\it i.e.}, a wino or
Higgsino. In such cases, the LSP has the electroweak interactions and
thus has a relatively large annihilation cross section, which is further
enhanced by the so-called Sommerfeld effects \cite{Hisano:2003ec,
*Hisano:2004ds}. Indeed, the thermal relic abundance of wino and
Higgsino with a mass of around 3~TeV \cite{Hisano:2006nn} and 1~TeV
\cite{Cirelli:2007xd}, respectively, is found to be in good agreement
with the observed dark matter density $\Omega_{\rm DM} h^2 \simeq 0.12$
\cite{Ade:2015xua}. Smaller masses are also allowed by the observation;
in such cases, their thermal relic accounts for only a part of the total
dark matter density and the rest should be filled with other dark
matter candidates and/or with non-thermal contribution via, {\it e.g.},
the late-time decay of gravitinos \cite{Gherghetta:1999sw,
*Moroi:1999zb}. For previous studies of wino and Higgsino dark matter,
see Refs.~\cite{Gherghetta:1999sw, *Moroi:1999zb, Hisano:2003ec,
*Hisano:2004ds, Hisano:2006nn, ArkaniHamed:2006mb, Shirai:2009fq,
*Ibe:2013jya, *Ibe:2014qya, Hisano:2015rsa, Hall:2012zp,
Ibe:2012sx, Cohen:2013ama, *Fan:2013faa, *Hryczuk:2014hpa,
Bhattacherjee:2014dya, Cirelli:2014dsa, Baumgart:2014saa,
Harigaya:2015yaa, Ibe:2015tma, *Hamaguchi:2015wga, Lefranc:2016fgn} and
Refs.~\cite{Nagata:2014wma, Cheung:2005pv, *Cheung:2009qk,
*Beylin:2009wz, *Jeong:2011sg, *Fox:2014moa, *Evans:2014pxa,
*Chun:2016cnm, *Dedes:2016odh}, respectively, and references therein. 

On the other hand, bino-like dark matter in general suffers from over-production, and thus a
certain mechanism is required to enhance the annihilation cross
section. For example, we may utilize the $s$-channel resonant
annihilation through the exchange of the Higgs bosons (called funnel)
\cite{Griest:1990kh}. Coannihilation \cite{Griest:1990kh} may
also work if there is a SUSY particle that is degenerate with the LSP in
mass and has a large annihilation cross section; stau
\cite{Ellis:1998kh, *Ellis:1999mm, *Nihei:2002sc, *Citron:2012fg}, stop
\cite{Boehm:1999bj, *Ellis:2001nx, *Ajaib:2011hs, *Ellis:2014ipa,
deSimone:2014pda, Liew:2016hqo}, gluino \cite{Profumo:2004wk,
*Feldman:2009zc, *Ellis:2015vaa, *Ellis:2015vna, deSimone:2014pda,
Harigaya:2014dwa,  Nagata:2015hha, Liew:2016hqo}, wino
\cite{Baer:2005jq, ArkaniHamed:2006mb, *Ibe:2013pua, Harigaya:2014dwa,
Nagata:2015pra}, {\it etc.}, can be such a candidate. In particular,
only the latter two can have a mass close to the LSP in the case of the
split-SUSY type models \cite{Wells:2003tf, *Wells:2004di,
ArkaniHamed:2004fb, *Giudice:2004tc,*ArkaniHamed:2004yi, Hall:2011jd,
Hall:2012zp, Ibe:2006de, Ibe:2011aa, *Ibe:2012hu, Arvanitaki:2012ps,
ArkaniHamed:2012gw, Evans:2013lpa, *Evans:2013dza, Nomura:2014asa}.
In this paper, we especially focus on the neutralino-gluino
coannihilation case as this turns out to offer a variety of
interesting signatures at colliders and thus can be probed in various
search channels at the LHC. For a search strategy of the bino-wino
coannihilation scenario at the LHC, see Ref.~\cite{Nagata:2015pra}.

In order for the neutralino-gluino coannihilation to work, the
neutralino LSP and gluino should be highly degenerate in mass. For
instance, if the neutralino LSP is bino-like, the mass difference
between the LSP and gluino, $\Delta m$, needs to be less than around
$100$~GeV and squark masses should be less than ${\cal O}(100)$~TeV for
coannihilation to be effective \cite{Nagata:2015hha}. Such a small mass
difference makes it difficult to probe this scenario in the conventional
LHC searches as hadronic jets from the decay products of gluinos tend to
be soft. In the previous work \cite{Nagata:2015hha}, however, it is
pointed out that such a compressed gluino with heavy squarks has a decay
length of $\gtrsim {\cal O}(1)$~mm and therefore may be probed by using
searches for displaced vertices (DVs). In fact, it was shown in
Ref.~\cite{Nagata:2015hha} that the DV search at the LHC can investigate
a wide range of parameter space where the correct dark matter abundance
is obtained for the bino LSP through coannihilation with
gluinos.

In this paper, we further study the prospects of the LHC searches to
probe this compressed gluino scenario. In particular, we discuss search
strategies for the very degenerate case, {\it i.e.}, $\Delta m \lesssim {\cal
O}(10)$~GeV. Such an extremely small mass difference considerably
narrows down the reach of DV searches. We however find that for such a
small mass difference, long-lived gluinos leave disappearing track
signals when they form charged $R$-hadrons, and thus can be probed in the
disappearing track searches. In addition, for gluinos which have a decay
length of $\gtrsim 1$~m, searches for anomalously large energy deposit
by (meta)stable heavy charged particles can be exploited. We see below that
these three searches are complementary to each other. Hence, by
combining the results from these searches we can thoroughly study the
compressed gluino scenario at the LHC.

%%%%%%%%%%%%%%%%%%%%%%%%%%%%%%%%%%%%%%%%%%%
\section{Properties of compressed gluino}
%%%%%%%%%%%%%%%%%%%%%%%%%%%%%%%%%%%%%%%%%%%%

Here we discuss the compressed gluino signatures at colliders. If
the masses of squarks are very large and/or the mass difference between the
gluino and the neutralino LSP $\Delta m$ is quite small, the gluino
decay width is strongly suppressed and its decay length can be as large
as the detector-size scale. We briefly discuss this feature in
Sec.~\ref{sec:gluinodecay}.\footnote{For detailed discussions on the
decay of (long-lived) gluinos, see Refs.~\cite{Haber:1983fc,
*Baer:1986au, *Barbieri:1987ed, *Baer:1990sc, *Toharia:2005gm,
*Gambino:2005eh, Sato:2012xf, *Sato:2013bta}.} When the gluino lifetime
is longer than the QCD hadronization time scale, gluinos produced at
colliders form $R$-hadrons. We discuss the properties of $R$-hadrons and their
implications on the LHC searches in
Sec.~\ref{sec:Rhadrons}.\footnote{For previous studies on the $R$-hadron
properties in the split-SUSY, see Ref.~\cite{Kilian:2004uj,
*Hewett:2004nw}. }

%%%%%%%%%%%%%%%%%%%%%%%%%%%%
\subsection{Gluino decay}
\label{sec:gluinodecay}
%%%%%%%%%%%%%%%%%%%%%%%%%%%%

%%%%%%%%%%%%%%%%%%%%%%%%%%%%%%%%%%%%%%%%%%%%%%%%%%%
\begin{figure}[t]
\centering
%\begin{comment}
\subcaptionbox{\label{fig:tree} Three-body
 decay}{\includegraphics[width=0.39\textwidth]{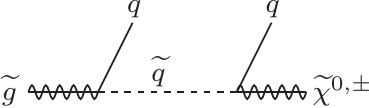}} 
\hspace{1cm}
\subcaptionbox{\label{fig:loop} Two-body
 decay}{\includegraphics[width=0.35\textwidth]{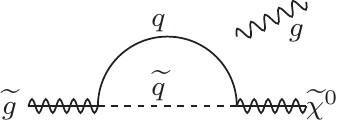}}  
%\end{comment}
\caption{
Diagrams that give rise to the gluino decay.
}
\label{fig:gluino_decay}
\end{figure}
%%%%%%%%%%%%%%%%%%%%%%%%%%%%%%%%%%%%%%%%%%%%%%%%%%%%%

The decay of gluinos is induced by the diagrams shown in
Fig.~\ref{fig:gluino_decay}. Here, we focus on the case where the mass
difference $\Delta m$ is rather small ($\Delta m \lesssim 100$~GeV) and
squark masses are much larger than the electroweak scale; {\it i.e.},
$\widetilde{m} \gtrsim 10$~TeV, where $\widetilde{m}$ denotes the
typical scale of squark masses.\footnote{We however note that gluinos
can be long-lived even though $\widetilde{m}$ is around the TeV scale if
the mass difference $\Delta m$ is small enough. This possibility is
briefly discussed in Sec.~\ref{sec:conclusion}.} In this 
case, the tree-level three-body decay process shown in
Fig.~\ref{fig:tree} is proportional to $(\Delta m)^5$, and thus its
decay width is strongly suppressed for compressed gluinos. 
For the loop-induced two-body decay in Fig.~\ref{fig:loop}, on the other
hand, the decay width strongly depends not only on the mass difference
but also on the nature of the lightest neutralino. If the neutralino LSP
is a pure bino and left-right mixing of the scalar top quarks is small,
the matrix element of the two-body decay process is  
suppressed by a factor of $m_{\widetilde g} - m_{\widetilde B}$
($m_{\widetilde g}$ and $m_{\widetilde B}$ are the gluino and bino
masses, respectively), which
originates from a chirality flipping in the external lines. As a result,
the two-body decay rate also goes as $\Gamma(\widetilde g \to \widetilde
B g) \propto (\Delta m)^5$, which makes the three-body decay channel dominate the two-body one. If the LSP is a pure wino, the
two-body decay is suppressed by the SU(2)$_L$ gauge symmetry, and thus
the three-body decay is again the dominant channel. Contrary to these
cases, if the LSP is Higgsino-like, the two-body decay is the main decay
process. In this case, the dominant contribution to this decay process
comes from the top-stop loop diagram and its matrix element
is proportional to the top mass $m_t$. Thus, the two-body decay rate
goes as $\Gamma(\widetilde g \to \widetilde H^0 g) \propto m_t^2 (\Delta
m)^3$. In addition, this loop contribution is logarithmically enhanced
as the masses of the scalar top quarks get larger. The tree-level decay
process is, on the other hand, suppressed by small Yukawa couplings
since the stop exchange process, which is large because of the top
Yukawa coupling, is kinematically forbidden for compressed
gluinos.

Eventually, we see that the gluino decay length is approximately given by
\begin{align}
c\tau_{\widetilde g} =
\begin{cases}
 {\cal O}(1-10)~{\rm cm} \times \left(\cfrac{\Delta m}{10~{\rm GeV}}\right)^{-5}
 \left(\cfrac{\widetilde m}{10~{\rm TeV}}\right)^{4}   & \text{(bino or wino LSP)}  \\[5pt]
 {\cal O}(0.01-0.1)~{\rm mm} \times  \left(\cfrac{\Delta m}{10~{\rm GeV}}\right)^{-3}
 \left(\cfrac{\widetilde m}{10~{\rm TeV}}\right)^{4}   & \text{(Higgsino LSP)}
 \end{cases}
~.
\label{eq:appgllftm}
\end{align}
The dependence of these approximated formulae on $\Delta m$ is captured
by the above discussions. From these expressions, we find that
compressed gluinos generically have decay lengths of the order of the
detector size when the squark mass scale is $\gtrsim 10$~TeV. We make
the most of this observation to probe the compressed gluino scenario.

There are several situations where the above tendency of gluino
decay may be altered. For the bino LSP case, the two-body decay channel can
be important if the left-right mixing in the scalar top sector is
large and $\Delta m$ is very small. This decay branch
can also be enhanced if there is a sizable bino-Higgsino mixing. In
addition, if $\Delta m \ll 10$~GeV, the parton-level description gets
less appropriate, and the hadronic properties of decay products
significantly affect the gluino decay. Further dedicated studies are
required to give a precise theoretical calculation of the gluino decay
rate for this very small $\Delta m$ region, which is beyond the scope of
this paper.

%%%%%%%%%%%%%%%%%%%%%%%%
\subsection{$R$-hadrons}
\label{sec:Rhadrons}
%%%%%%%%%%%%%%%%%%%%%%%%

A long-lived gluino forms a bound state with quarks and/or gluons once
they are produced at colliders. Such bound states, being $R$-parity odd,
are called $R$-hadrons \cite{Farrar:1978xj}. $R$-hadrons are categorized into
several classes in terms of their constituents; if $R$-hadrons are
composed of a gluino and a pair of quark and anti-quark, $\widetilde{g}
\bar{q}q$, they are called $R$-mesons; if they consist of a gluino and
three quarks (anti-quarks), $\widetilde{g} qqq$ ($\widetilde{g}
\bar{q}\bar{q} \bar{q}$), they are called $R$-baryons ($R$-antibaryons);
a bound state which is made of a gluino and a gluon, $\widetilde{g} g$, is
referred to as an $R$-glueball.

The production fractions of $R$-hadron species have a direct impact on
the gluino search sensitivities discussed below, as some of these
search strategies rely on the production of charged $R$-hadrons. A
computation \cite{Fairbairn:2006gg} in which hadronization is performed
using {\sc Pythia} \cite{Sjostrand:2006za} shows that the production
rates of $R$-mesons dominate those of $R$-baryons. The production
fraction of $R$-glueball is, on the other hand, theoretically unknown
and thus regarded as a free parameter. In the analysis of
Ref.~\cite{Fairbairn:2006gg}, this fraction is set to be 10\%, which is
the default value used in {\sc Pythia}. Then, it is found that the
fraction of $R$-mesons is 88.5\% while that of $R$-baryons is only
1.6\%. Among them, charged $R$-hadrons are 44.8\%. This value
however significantly decreases if we set the $R$-glueball fraction to
be a larger value. Taking this ambiguity into account, in the following
analysis, we take different values for the $R$-glueball fraction and
regard the resultant changes as theoretical uncertainty.

The mass spectrum of $R$-hadrons, especially that of $R$-mesons and
$R$-glueball, also affects the $R$-hadron search strategy
significantly. An estimation \cite{Kraan:2004tz} based on a simple mass
formula for the lowest hadronic states \cite{DeRujula:1975qlm}, which is
derived from the color-spin interaction given by one gluon exchange,
indicates that the lightest $R$-meson state is ``$R$-rho'', namely, a bound
state which consists of a gluino and a vector iso-triplet made of up and
down quarks. The mass of ``$R$-pion'' (a bound state of a gluino and a
spin-zero iso-triplet made of up and down quarks) is found to be larger
than the $R$-rho mass by about 80~MeV. This observation is consistent
with the calculations using the bag model \cite{Chanowitz:1983ci} and
QCD lattice simulation \cite{Foster:1998wu}, which predict $R$-rho to be
lighter than $R$-pion by 40~MeV and 50~MeV, respectively. On the other
hand, there is controversy about the estimation of the $R$-glueball
mass. An estimation by means of the constituent masses of partons
shows that $R$-glueball is heavier than $R$-rho by 120~MeV
\cite{Fairbairn:2006gg}. The bag-model calculation
\cite{Chanowitz:1983ci} also predicts $R$-glueball to be slightly heavier
than $R$-rho. However, the lattice result \cite{Foster:1998wu} shows that
the $R$-rho mass is larger than the $R$-glueball mass by 47~MeV, though
we cannot conclude by this result that these results are incompatible
with each other since the uncertainty of this calculation is as large as
90~MeV (and the former two estimations also suffer from uncertainties of
similar size). We here note that if $R$-glueball is lighter than
$R$-rho and the mass difference between them is larger than the pion
mass, then an $R$-rho can decay into an $R$-glueball and a pion via
strong interactions. Other $R$-mesons may also decay into $R$-glueball.
This considerably reduces the number of tracks associated with charged
$R$-mesons, and thus weakens the discovery reach of $R$-hadrons. In the
following analysis, we assume that such decay channels are kinematically
forbidden, as is supported by the above calculations of the $R$-meson
and $R$-glueball mass spectrum. As for
$R$-baryons, the flavor singlet $J=0$ state, which has a non-zero
strangeness, is the lightest \cite{Farrar:1984gk,
Buccella:1985cs}. In addition, there are flavor octet states which are
stable against strong decays and heavier than the singlet state
by about a few hundred MeV. Their weak-decay lifetime is likely to be
sufficiently long so that they can be regarded as stable at colliders
\cite{Farrar:2010ps}.

While $R$-hadrons are propagating through a detector, they may scatter
off nuclei in it. Such processes are potentially important since they may change $R$-hadron species. For instance, by scattering a nucleon in the detector
material, an $R$-meson or $R$-glueball can be converted into an
$R$-baryon with emitting a pion. 
However, the reverse process is unlikely since pions
are rarely found in the detector material and the process itself suffers
from kinematical suppression. For this reason, although $R$-mesons and
$R$-glueball are dominantly produced at the outset, we have a sizable
fraction for $R$-baryons in the outer part of detectors, such as in the
Muon Spectrometer. The nuclear reaction rates of $R$-hadrons are
evaluated in Refs.~\cite{Kraan:2004tz, Mackeprang:2006gx,
Mackeprang:2009ad, Farrar:2010ps}, and it is found that an $R$-hadron
interacts with nucleons about five times while propagating in 2~m of
iron. Therefore, most of $R$-mesons and $R$-glueballs may be converted into
$R$-baryons before they enter into the Muon Spectrometer. In the
analysis discussed below, however, we focus on the $R$-hadron searches
using the Inner Detector, on which the nucleon interactions have little
impact as the matter density up to the Inner Detector region is
very low. This makes our search strategies free from uncertainties
originating from the estimation of $R$-hadron interactions in
detectors---we neglect these effects in the following analysis.

%%%%%%%%%%%%%%%%%%%%%%%%%%%%%%%%%%%%%%%%%%%%%%%
\section{LHC searches}
\label{sec:results}
%%%%%%%%%%%%%%%%%%%%%%%%%%%%%%%%%%%%%%%%%%%%%%%

Next, we discuss the LHC signatures of the compressed gluinos. Depending
on the gluino lifetime and the gluino-LSP mass difference, we need
adopt several strategies to catch gluino signals. In this paper, we
focus on the ATLAS detector. The performance of the CMS detector is
similar to that of the ATLAS, though.

%%%%%%%%%%%%%%%%%%%%%%%%%%%%%%%%
\subsection{ATLAS experiment}
\label{sec:atlasex}
%%%%%%%%%%%%%%%%%%%%%%%%%%%%%%%%

The ATLAS detector is located at one of the interaction points of the LHC,
which consists of the Inner Detector, the calorimeters, the Muon
Spectrometer, and the magnet systems \cite{PERF-2007-01}. The long-lived
gluino searches discussed in this paper make full use of the Pixel
detector and the SemiConductor Tracker (SCT) in the Inner
Detector. Various dedicated techniques, which have been developed to
search for new long-lived particles, can be applied to the long-lived
gluino searches in order to maximize its discovery potential. Here, we
briefly review the detectors used in the searches relevant to this work.

%%%%%%%%%%%%%%%%%%%%%%%%%%%%%%%%%%%
\subsubsection*{Pixel detector}
%%%%%%%%%%%%%%%%%%%%%%%%%%%%%%%%%%%

The Pixel detector is the sub-detector closest to the interaction point,
which has a four-layer cylindrical structure with a length of about
$800~{\rm mm}$ in the barrel region. The innermost layer called
Insertable B-Layer (IBL), which was installed before the LHC-Run2
started \cite{Capeans:1291633}, has silicon pixel sensors of
$50\times250~{\rm \mu m}$ and is located at a radius of $33~{\rm
mm}$. The other layers have silicon pixel sensors of $50\times400~{\rm
\mu m}$ at radii of $50.5~\text{mm}$, $88.5~\text{mm}$, and
$122.5~\text{mm}$. The Pixel detector can measure the energy deposit
along the trajectory of each charged particle, {\it i.e.}, $dE/dx$,
which is sensitive to slow-moving (meta)stable particles according to
the Bethe-Bloch formula.

%%%%%%%%%%%%%%%%%%%%%%
\subsubsection*{SCT}
%%%%%%%%%%%%%%%%%%%%%%

The SCT surrounds the Pixel detector, which has four layers in the
barrel region with a length of about $1500~\text{mm}$ at radii of
$299~\text{mm}$, $371~\text{mm}$, $443~\text{mm}$, and
$514~\text{mm}$. A module on each layer consists of two $80~\mu {\rm
m}$-pitch strip silicon sensors with a stereo angle of $40~\text{mrad}$
between strip directions. The SCT does not have a capability to measure
$dE/dx$ in contrast to the Pixel detector, since the SCT employs binary
readout architecture.

%%%%%%%%%%%%%%%%%%%%%%%%%%%%%%%%%%%%
\subsection{ATLAS searches}
\label{sec:glusig}
%%%%%%%%%%%%%%%%%%%%%%%%%%%%%%%%%%%%

Gluinos give rise to various signatures at colliders depending on the
decay lifetime. In this work, we consider the following four search
strategies,\footnote{Some of these search strategies have already been
considered in the context of long-lived gluino searches in
Refs.~\cite{Nagata:2015hha, Liu:2015bma}.} which are sensitive to
different ranges of the gluino decay length: 

%%%%%%%%%%%%%%%%%%%%%%%%%%%%%%%%
\subsubsection*{Prompt decay \cite{ATLAS-CONF-2016-078}:
   $c\tau_{\widetilde g} \lesssim 1~{\rm mm}$} 
%%%%%%%%%%%%%%%%%%%%%%%%%%%%%%%%%

Searches for a new particle which is assumed to decay at the interaction
point are also sensitive to long-lived particles. The ATLAS experiment
reconstructs tracks whose transverse impact parameters ($d_0$) are less
than $10~\rm{mm}$, then checks a correspondence with primary vertices.
As a result, a portion of the decay vertices of long-lived particles is
merged with one of the primary vertices.

Generally, for metastable gluinos, these inclusive searches get less
effective for $c\tau_{\widetilde g} \gg 1$~cm, since jets from the
gluino decay are displaced from the primary vertex and fail the event
selection criteria \cite{ATLAS:2014qga}. In the compressed gluino case,
however, these searches are less sensitive to the gluino lifetime, since
it is jets from the initial state radiation that play a main role in the
conventional jets + MET searches. For this reason, even for a gluino
with a decay length greater than ${\cal O}(1)$~m, the resultant mass
bound will be similar to that for a prompt decay compressed gluino.

%%%%%%%%%%%%%%%%%%%%%%%%%%%%%%%%%%%%%%%%%%%%%%%%%%%%%%%%%%%%
\subsubsection*{Displaced-vertex search \cite{Aad:2015rba}:
   $c\tau_{\widetilde g} \gtrsim 1~{\rm mm}$} 
%%%%%%%%%%%%%%%%%%%%%%%%%%%%%%%%%%%%%%%%%%%%%%%%%%%%%%%%%%%%

A long-lived gluino decaying to quarks or a gluon leaves a displaced
vertex (DV) away from the interaction point. In order to reconstruct tracks
from such a DV, the requirement on the transverse impact parameter for
tracks is loosen such that $2~{\rm mm} < |d_0| < 300~{\rm mm}$. As a
result, the sensitivity is maximized for particles with a decay length
of ${\cal O}(10)~{\rm mm}$. 

The sensitivity becomes worse as the mass difference between gluino and
the LSP gets smaller \cite{Nagata:2015hha}, since the invariant mass of
DVs is required to be larger than $10~\text{GeV}$ in order to separate
signal events from background fake vertices. Due to this requirement,
the gluino-LSP mass difference of $\lesssim 20$~GeV is hard to probe.

%%%%%%%%%%%%%%%%%%%%%%%%%%%%%%%%%%%%%%%%%%%
\subsubsection*{Disappearing-track search \cite{SUSY-2013-01}:
   $c\tau_{\widetilde g} \gtrsim 10~{\rm cm}$} 
%%%%%%%%%%%%%%%%%%%%%%%%%%%%%%%%%%%%%%%%%%%

Originally, the disappearing-track search has been developed for the
search of long-lived charged winos with the neutral wino being the
LSP \cite{Feng:1999fu, Ibe:2006de, Asai:2007sw, *Asai:2008sk,
*Asai:2008im}. This technique may also be applicable to long-lived
gluinos for the 
following reason. As we discuss in the previous section, a certain
fraction of long-lived gluinos form charged $R$-hadrons. If the gluino and
the LSP are degenerate in mass, the track associated with a charged
$R$-hadron seems to disappear when the gluino in the $R$-hadron decays,
since the jet emission from the gluino decay is very soft. As mentioned
above, the DV search does not work efficiently when the gluino-LSP mass
difference is very small. Such a degenerate mass region can instead be
covered by the disappearing-track search.

A candidate track in the disappearing-track search should have four hits
in the Pixel detector and the SCT with no activity after the last hit 
required. Thanks to the installation of IBL, the minimum length of
disappearing tracks which can be searched for by this strategy is
shorten from 299~mm to 122.5~mm in the LHC Run2. This allows the range
of decay lengths covered by the disappearing-track search to be slightly
wider than that in the DV search. 

%%%%%%%%%%%%%%%%%%%%%%%%%%%%%%%%%%%%%%%%%%%%%%%%%%%%%%%%%%%%%
\subsubsection*{Pixel $dE/dx$ search \cite{SUSY-2016-03}:
   $c\tau_{\widetilde g} \gtrsim 1~{\rm m}$}  
%%%%%%%%%%%%%%%%%%%%%%%%%%%%%%%%%%%%%%%%

A particle with a mass of the order of the electroweak scale or larger
tends to travel with a low velocity after it is produced at
the LHC, which may be observed as a large $dE/dx$ in the Pixel
detector. Hence, by searching for this signature, we can probe charged
$R$-hadrons. While a minimum ionization particle is expected to give $\sim
1.2~\text{MeV} \cdot {\rm cm}^2 /{\rm g}$ of $dE/dx$, the threshold for
the Pixel $dE/dx$ search is set to be $1.8~\text{MeV} \cdot \text{cm}^2
/\text{g}$ with a small correction depending on $\eta$.
For the track selection, at least seven hits in the Pixel detector and
the SCT are required, which corresponds to the minimum track-length of
$371~{\rm mm}$. Note that this search strategy does not require decay of
particles, and thus is also sensitive to completely stable
particles. For this reason, the cover-range of decay lengths by the
Pixel $dE/dx$ search is quite broad. 

%%%%%%%%%%%%%%%%%%%%%%%%%%%%%%%%%%%%%%%
\subsection{LHC Prospects}
%%%%%%%%%%%%%%%%%%%%%%%%%%%%%%%%%%%%%%%

%%%%%%%%%%%%%%%%%%%%%%%%%%%%%%%%
\begin{figure}[t]
\centering
%\begin{comment}
\includegraphics{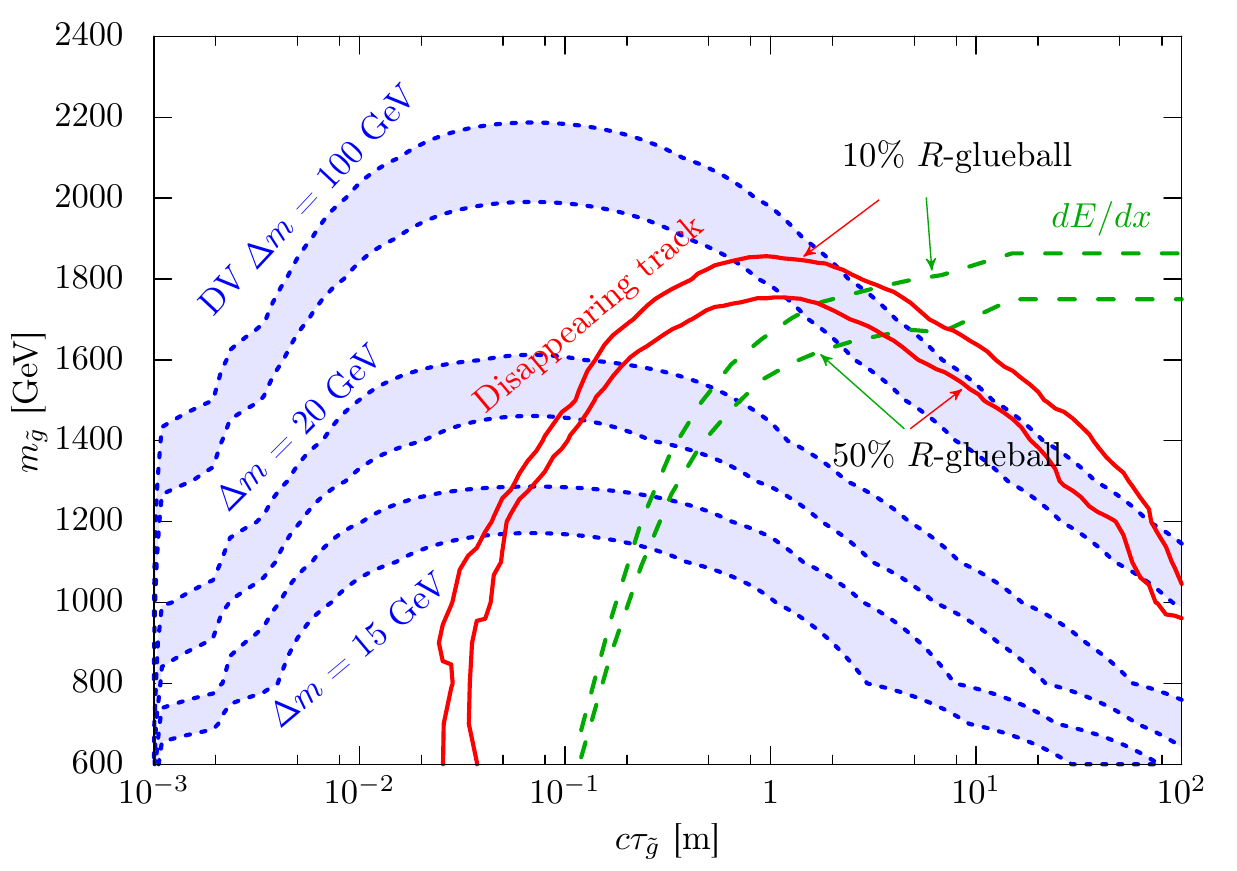} 
%\end{comment}
\caption{
 The expected limits from the searches considered in Sec.~\ref{sec:glusig}
 with an integrated luminosity of 40~fb$^{-1}$ at the 13~TeV LHC. The
 blue dashed, red solid, and green long-dashed lines correspond to the
 DV search, the disappearing-track search, and the Pixel
 $dE/dx$ search, respectively. 
}
\label{fig:gluino_decay_constraint}
\end{figure}
%%%%%%%%%%%%%%%%%%%%%%%%%%%%%%%%

Now, we show the sensitivities of the searches listed in
Sec.~\ref{sec:glusig} to the compressed gluino scenario.  
In this study, we use the program {\sc Madgraph5}
\cite{Alwall:2011uj}$+${\sc Pythia8}
\cite{Sjostrand:2006za,Sjostrand:2007gs}$+${\sc Delphes3}
\cite{deFavereau:2013fsa} and estimate the production cross sections of
SUSY particles with {\sc Prospino2} \cite{Beenakker:1996ed} or {\sc
NLL-fast}~\cite{Beenakker:1996ch,*Kulesza:2008jb,*Kulesza:2009kq,
*Beenakker:2009ha,*Beenakker:2011fu}.

In Fig.~\ref{fig:gluino_decay_constraint}, we show the prospects of each
search in the $c \tau_{\tilde{g}}$--$m_{\tilde{g}}$ plane with an
integrated luminosity of 40~fb$^{-1}$ at the 13~TeV LHC. The blue dashed
lines show the expected limits from the DV search with the
gluino-LSP mass difference set to be $\Delta m =100$~GeV, 20~GeV, and
15~GeV from top to bottom. We use the same event selection requirements
as in the 8 TeV study \cite{Nagata:2015pra} except that we require
missing energy $E_{\rm T}^{\rm miss}$ to be greater than 200~GeV as a
trigger, which was $E_{\rm T}^{\rm miss} > 100$~GeV in the 8 TeV study
\cite{Nagata:2015pra}. We expect the number of background events for
this signal is as small as in the case of the 8 TeV run
\cite{Aad:2015rba}; here, we assume it to be 0--10 and the systematic
uncertainty of the background estimation to be 10\%. The upper (lower)
border of each band corresponds to the case where the number of the
background events is 0 (10). It is found that the sensitivity of the DV
search is maximized for a gluino with a decay length of $\sim 10$~cm,
and may reach a gluino mass of about 2.2~TeV (1.2~TeV) for $\Delta m =
100$~GeV (15~GeV). We also find that the DV search becomes less powerful
when the gluino-LSP mass difference gets smaller, as mentioned in
Sec.~\ref{sec:glusig}.

The red dashed lines show the expected reaches of the disappearing-track
search. As we discussed above, this search relies on the production of
charged $R$-hadrons, but the estimate of the charged $R$-hadron
production rates suffers from uncertainty due to the unknown
$R$-glueball fraction. To take this uncertainty into account, in
Fig.~\ref{fig:gluino_decay_constraint}, we set the $R$-glueball fraction
to be 10\% and 50\% in the top and bottom lines, respectively. To estimate the
prospects, we basically adopt the same selection criteria as the charged
wino search at the LHC Run 1 \cite{Aad:2013yna}, including the isolation
criteria and the detection efficiency of a disappearing track. To consider
the update in the LHC Run 2, we refer to the result presented in
Ref.~\cite{Saito:jsps}. In addition, to take into account the
improvement because of the IBL, we assume the detection efficiency 60\% for
$|\eta|<1.5$ and $13<r<30$~cm, which was zero at the Run 1 study
\cite{Aad:2013yna}. We further adopt the following kinematic cut:
$E_{\rm T}^{\rm miss} > 140$~GeV and the leading jet with a transverse
momentum of $P_{\rm T}>140$~GeV. The number of background events is
assumed to be 10 and its uncertainty is supposed to be 10\% for
40~fb$^{-1}$. The result in Fig.~\ref{fig:gluino_decay_constraint} shows
that the disappearing track search is sensitive to a 
decay length of $\sim 1$~m, and can probe a longer lifetime region than
the DV search. Notice that even for $c \tau_{\tilde{g}} =
{\cal O}(10)$~cm, around which the sensitivity of the DV search is
maximized, the reach of the disappearing-track search may exceed that of
the DV search if the gluino-LSP mass difference is very small.

The expected reach of the Pixel $dE/dx$ search is plotted in the green
long-dashed lines. Here again, we set the $R$-glueball fraction to be
10\% and 50\% in the top and bottom lines, respectively, to show the
uncertainty from the unknown $R$-glueball fraction. For this analysis,
we adopt the same event selection as in the study of ATLAS with
$\sqrt{s}=13$~TeV and an integrated luminosity of 3.2~${\rm fb}^{-1}$
\cite{Aaboud:2016dgf}, where the $R$-glueball fraction is set to be
10\%. We have estimated the number of background events by rescaling the
result of 3.2~${\rm fb}^{-1}$ to 40~${\rm fb}^{-1}$. We have also
assumed that the gluino-LSP mass difference is so tiny that the missing
energy, which is required in Ref.~\cite{Aaboud:2016dgf}, comes from
initial state radiations. As can be seen, the Pixel $dE/dx$ search can
probe $c \tau_{\tilde{g}} \gtrsim 1$~m, and its sensitivity does not
decrease for $c \tau_{\tilde{g}} \gg 1$~m, contrary to the previous two
cases.

%%%%%%%%%%%%%%%%%%%%%%%%%%%%%%%%%%%%%%%%%%%%%%%%%%%
\begin{figure}[t]
\centering
%\begin{comment}
\subcaptionbox{\label{fig:10tev} $m_{\tilde{q}} =
 10$~TeV}{\includegraphics[width=0.49\textwidth]{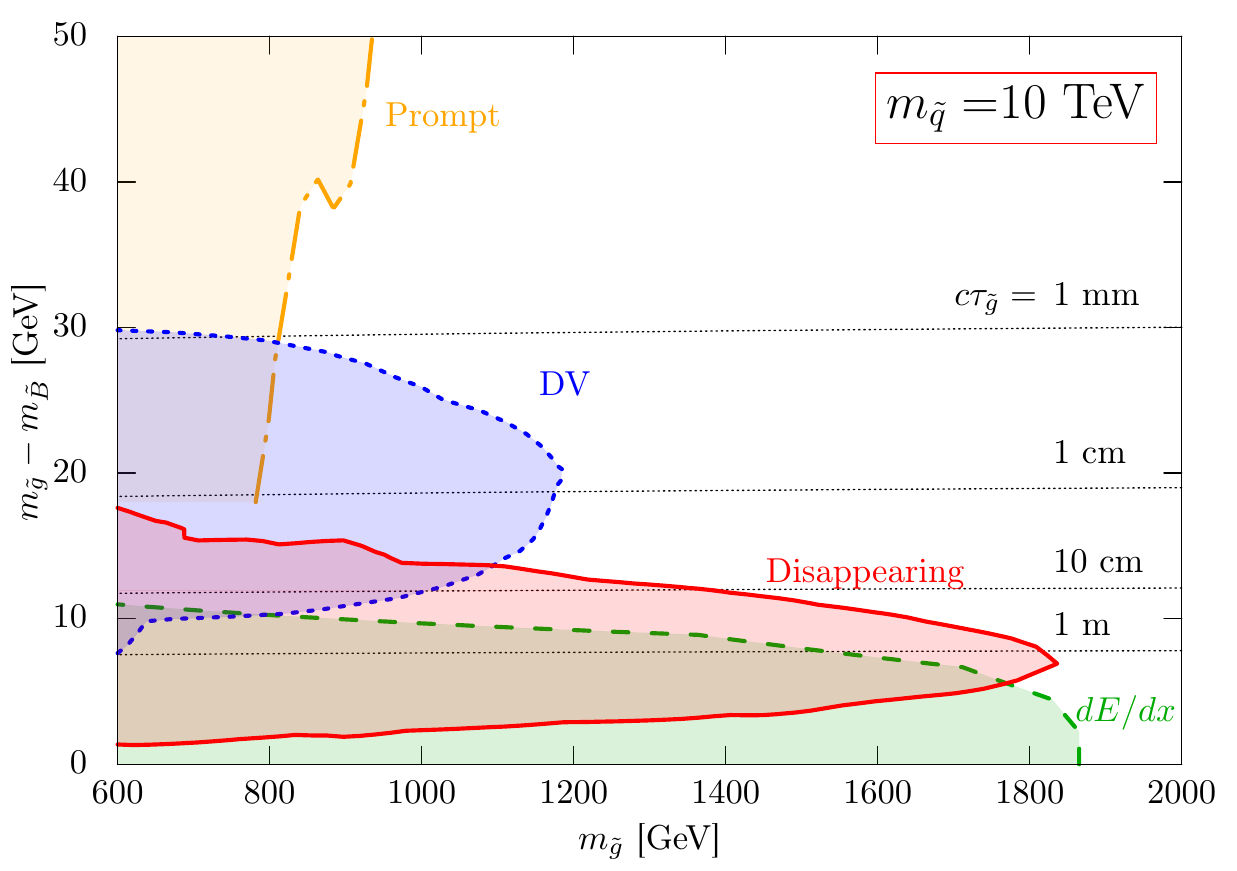}} 
%\hspace{1cm}
\subcaptionbox{\label{fig:50tev} $m_{\tilde{q}} =
 50$~TeV}{\includegraphics[width=0.49\textwidth]{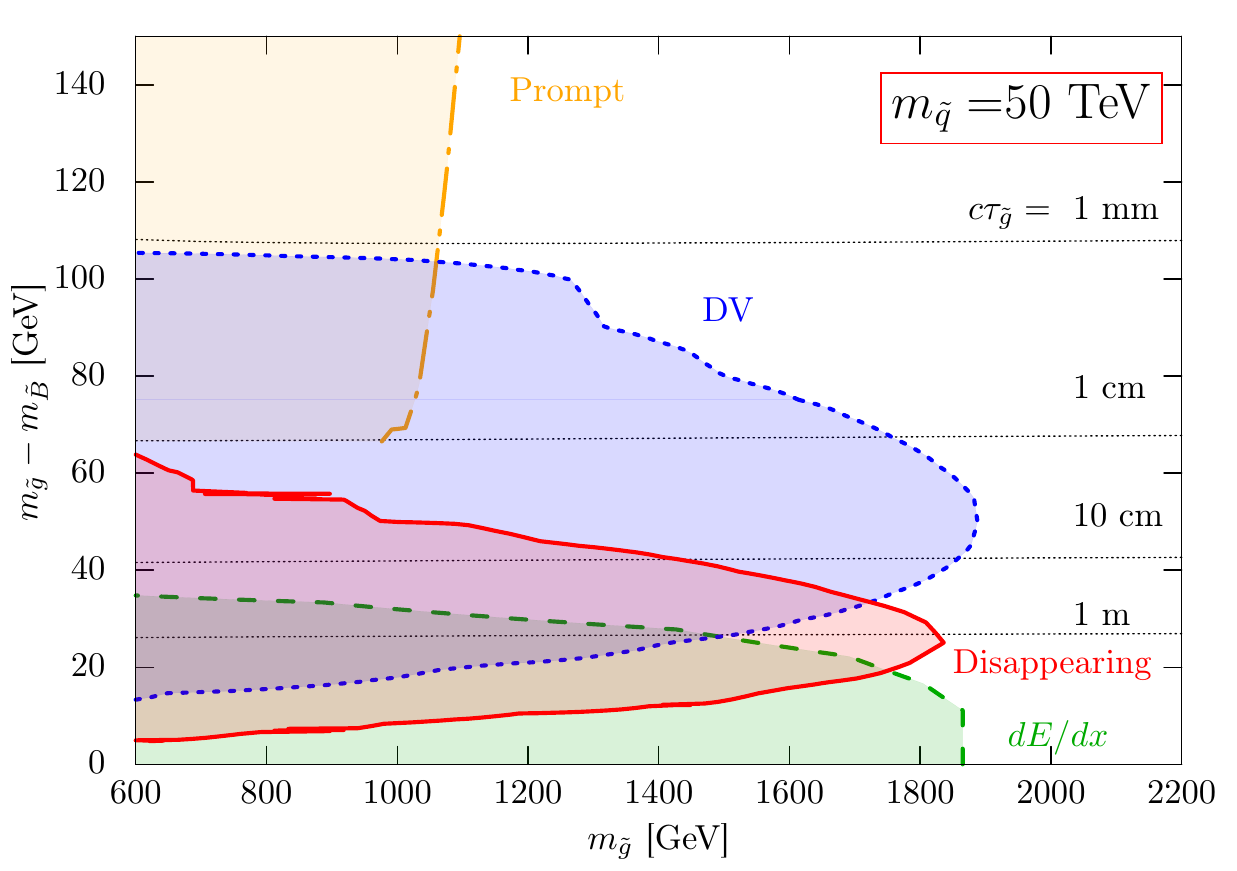}}  
%\end{comment}
\caption{ The expected limits on the compressed gluino parameter space from the
 13~TeV LHC with an integrated luminosity of 40~fb$^{-1}$. The shaded
 areas with the blue dashed, red solid, green long-dashed, and orange
 dash-dotted borders can be probed with the searches of 
 DVs, disappearing tracks, anomalous energy deposit ($dE/dx$) in
 the Pixel detector, and prompt gluino decays, respectively. The black
 dotted lines show a contour plot for the gluino decay length
 $c\tau_{\tilde{g}}$. 
}
\label{fig:massdiff}
\end{figure}
%%%%%%%%%%%%%%%%%%%%%%%%%%%%%%%%%%%%%%%%%%%%%%%%%%%%%

To show the prospects of the above searches for probing the compressed
gluino scenario, in Fig.~\ref{fig:massdiff}, we show their expected
reaches with the 13~TeV 40~fb$^{-1}$ LHC data in the
$m_{\tilde{g}}$--$\Delta m$ plane, where the shaded areas with the blue
dashed, red solid, green long-dashed, and orange dash-dotted borders
correspond to the searches of DVs, disappearing tracks,
anomalous energy deposit ($dE/dx$) in the Pixel detector, and prompt
gluino decays, respectively. We also show a contour plot for the gluino
decay length $c\tau_{\tilde{g}}$ in the black dotted lines. All squark
masses (collectively denoted by $m_{\tilde{q}}$) are set to be 10~TeV
and 50~TeV in Figs.~\ref{fig:10tev} and \ref{fig:50tev},
respectively, and the LSP is assumed to be a pure bino. To obtain the
expected sensitivity of the prompt gluino decay search with the
40~fb$^{-1}$ data, we adopt the event cut criteria used in
Ref.~\cite{ATLAS-CONF-2016-078} and estimate the number of background
events by rescaling the result of the 13.3~fb$^{-1}$ case. 
These figures show that the highly compressed region can
be probed with the disappearing-track search and the Pixel $dE/dx$
search, while if the gluino-LSP mass splitting is large enough (such
that $c\tau_{\tilde{g}} \ll 1$~mm), only the traditional gluino searches
can have sensitivities. Between these two regions, the DV
search offers the best sensitivity. We also find that the reach of the
DV search strongly depends on the squark masses.
To obtain a gluino decay length of $c\tau_{\tilde{g}} \sim
10$~cm, to which the DV search is most sensitive, a
smaller $\Delta m$ is required for lighter squarks (see
Eq.~\eqref{eq:appgllftm}). As noted above, the sensitivity of the
DV search is considerably reduced for a small $\Delta m$;
for this reason, the reach of the DV search shrinks for
small squark masses. On the other hand, the disappearing-track and Pixel
$dE/dx$ searches are rather robust on the change of $m_{\tilde{q}}$, as
these searches do not rely on the jet emission from the gluino decay. In
any case, Fig.~\ref{fig:massdiff} shows that the experimental strategies
discussed in Sec.~\ref{sec:glusig} play complementary rolls in searching for
long-lived gluinos, and by combining the results from these searches we
can probe a wide range of parameter space in the compressed gluino
scenario.

%%%%%%%%%%%%%%%%%%%%%%%%%%%%%%%%%%%%%%%%
\section{Conclusion and discussion}
\label{sec:conclusion}
%%%%%%%%%%%%%%%%%%%%%%%%%%%%%%%%%%%%%%%%

\begin{figure}[t]
\centering
%\begin{comment}
\includegraphics{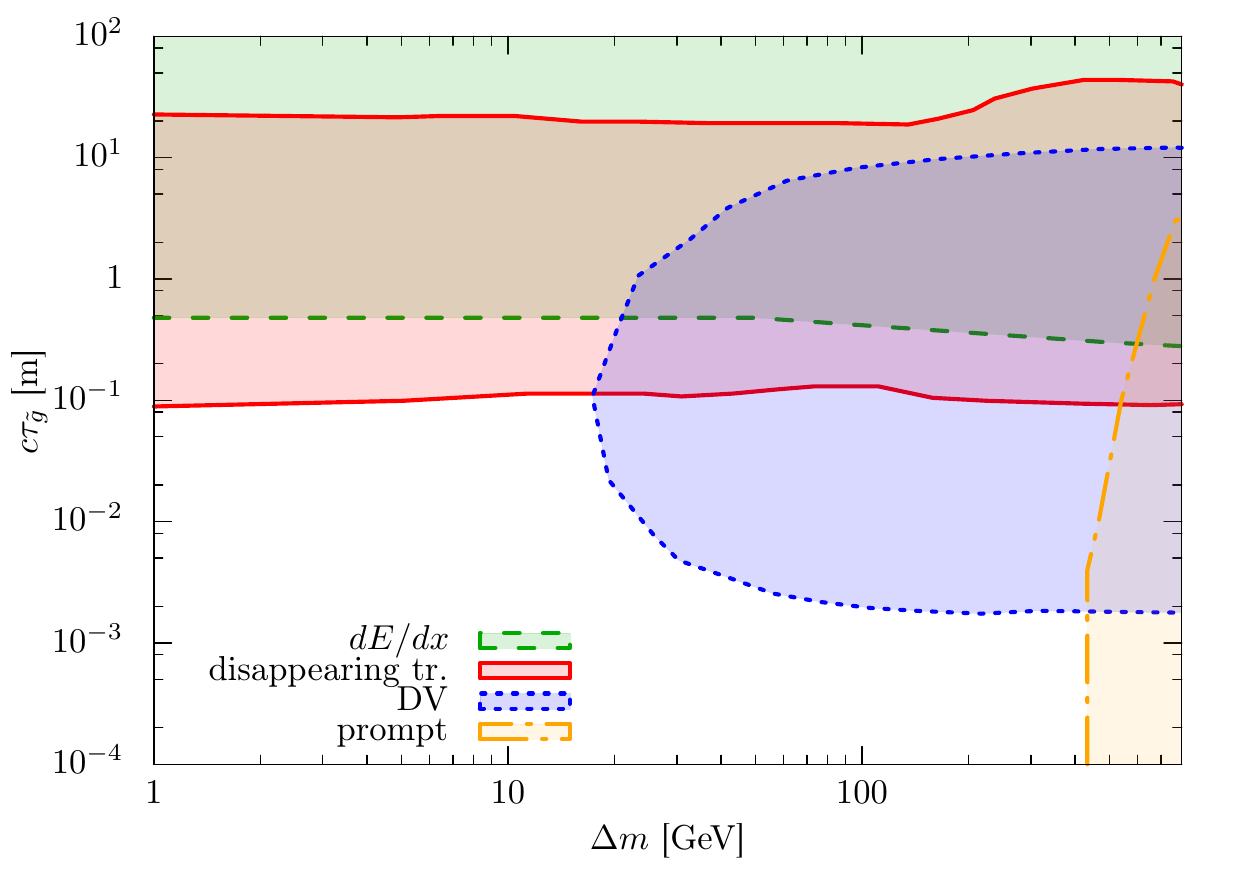} 
%\end{comment}
\caption{
Testable regions of the gluino-LSP mass difference $\Delta m$ and the
 gluino decay length $c\tau_{\widetilde{g}}$ from the $dE/dx$,
 disappearing track, displaced vertex, and prompt inclusive searches at
 13~TeV LHC with an integrated luminosity of 40~fb$^{-1}$. Here we fix
 the gluino mass to be 1.5~TeV.
}
\label{fig:g1500}
\end{figure}

The compressed-gluino parameter region in the MSSM, where gluino and the
LSP are highly degenerate in mass, can evade the over-abundance of dark
matter and thus is still viable. Therefore, it is important to test this
possibility experimentally. It is however difficult to probe this
scenario using the conventional search strategy at the LHC based on the
jets plus missing energy signatures, since jets emitted from the gluino
decay tend to be very soft. In this paper, we have discussed the
LHC search strategies which are sensitive to the compressed gluino
scenario. They include the searches of DVs, disappearing
tracks, and anomalous energy deposit in the Pixel detector, on top of
the ordinary inclusive searches. Then we have found that these searches
are indeed sensitive to long-lived gluinos.

In summary, we show the cover areas of these search strategies in the
$\Delta m$--$c\tau_{\widetilde{g}}$ plane in Fig.~\ref{fig:g1500}, where
we set the gluino mass to be 1.5~TeV and consider the 13~TeV LHC run
with an integrated luminosity of 40~fb$^{-1}$. As seen in this figure,
depending on the lifetime and the gluino-LSP mass difference, we can
adopt different search strategies. When the gluino decay length is
$\gtrsim 1$~m, the Pixel $dE/dx$ search offers the best sensitivity to
gluinos. For $0.1~{\rm m} \lesssim c\tau_{\tilde{g}} \lesssim 10~{\rm m}$, the disappearing-track
search is quite promising. The DV search can cover the
range of $1~{\rm mm} \lesssim c \tau_{\tilde{g}} \lesssim 1$~m, though
its sensitivity strongly depends on the gluino-LSP mass difference. The 
$c \tau_{\tilde{g}} < 1$~mm region can be probed by ordinary
prompt-decay gluino searches. As a consequence, these searches
complement each other, which allows us to investigate a broad range of
the compressed gluino region in the LHC experiments.

Notice that although we have focused on the cases with a high
sfermion mass scale ($\widetilde{m} \gtrsim 10$~TeV), the search
strategies discussed in this paper, especially the disappearing-track
search and the $dE/dx$ search, can also be powerful for
$\widetilde{m} < 10$~TeV if the gluino-LSP mass difference is small
enough to make gluino long-lived. Such a possibility may be
interesting since it offers a refuge for (semi)natural SUSY models, which
may be uncovered by the long-lived gluino searches.

In this paper, we discuss long-lived gluinos whose longevity
comes from small mass difference between the gluino and the
LSP. Actually, in not only the gluino case but more general ``co-LSP''
scenarios, we may have such a long-lived particle. For instance, in the
very compressed stau-LSP and stop-LSP cases, metastable charged
particles may appear and thus can be observed in the long-lived particle
searches at the LHC. As it turns out, for
instance, by using the setup of the disappearing track search discussed
above, we can probe a right-handed stau (stop) with a mass of around 200
(950)~GeV at the LHC Run 2 for $c\tau=10$~cm. Detailed studies of such
generalization are out of the scope of the present paper and will be
discussed elsewhere.

%%%%%%%%%%%%%%%%%%%%%%%%%%%%%%%%%%%%
\section*{Acknowledgments}
%%%%%%%%%%%%%%%%%%%%%%%%%%%%%%%%%%%%

This work was supported by World Premier International Research Center
Initiative (WPI), MEXT, Japan.

%%%%%%%%%%%%%%%%%%%%%%%%%%%%%%%%%%%%%%%%%%%%%%
%\section*{Appendix}
%\appendix
%%%%%%%%%%%%%%%%%%%%%%%%%%%%%%%%%%%%%%%%%%%%%

%%%%%%%%%%%%% References %%%%%%%%%%%%%%%%%%%
\bibliographystyle{aps}
\bibliography{ref}
%%%%%%%%%%%%%%%%%%%%%%%%%%%%%%%%%%%%%%%%%%%%

\end{document}